 Latex with RevTex, 3 Postscript figures (not included)
 available on request.

\documentstyle[preprint,revtex]{aps}
\tightenlines
\begin{document}
\draft
\begin{title}
The Momentum Dependence of the $\rho-\omega$ Mixing\\
Amplitude in a Hadronic Model.
\end{title}
\author{J. Piekarewicz}
\begin{instit}
Supercomputer Computations Research Institute, \\
Florida State University, Tallahassee, FL 32306
\end{instit}
\author{A.G.~Williams}
\begin{instit}
Department of Physics and
Supercomputer Computations Research Institute, \\
Florida State University, Tallahassee, FL 32306
\end{instit}

\begin{abstract}
We calculate the momentum dependence of the $\rho-\omega$
mixing amplitude in a purely hadronic model. The basic
assumption of the model is that the mixing amplitude is
generated by $N{\bar{N}}$ loops and thus driven entirely
by the neutron-proton mass difference. The value of the
amplitude at the $\omega$-meson point is expressed in terms
of only the $NN\omega$ and the $NN\rho$ coupling constants.
Using values for these couplings constrained by empirical
two-nucleon data we obtain a value for the mixing amplitude in
agreement with experiment. Extending these results to the
spacelike region, we find a $\rho-\omega$ contribution to the
NN interaction that is strongly suppressed and opposite in sign
relative to the conventional contribution obtained from using
the constant on-shell value for the mixing amplitude.
\end{abstract}

\narrowtext

      The existence of a nonzero mixing matrix element between the
isovector $\rho$ meson and the isoscalar $\omega$ meson is by now
firmly established~\cite{barkov85,henmil79,miller90}.
In addition to the observed (small) branching
ratio for the G-parity forbidden decay of the $\omega$ meson into
two pions, the understanding of the pion form factor at the
$\omega$-meson point ($q^{2}=m_{\omega}^{2})$ necessitates the coherent
interplay of two distinct amplitudes; a dominant, G-parity allowed
contribution
($\gamma\!\!~\rightarrow\!\!~\rho\!\!~\rightarrow\!\!~2\pi$) interfering
with a small ($\gamma\!\!~\rightarrow\!\!~\omega\!\!~\rightarrow\!\!
{}~\rho\!\!~\rightarrow~2\pi$) amplitude
arising from $\rho-\omega$ mixing~\cite{barkov85,miller90}.

      It has long been recognized that a $\rho-\omega$ mixing amplitude
would give rise to charge symmetry violation (CSV) in the nucleon-nucleon
(NN) force~\cite{henmil79,coon75,coon77}.
The $\rho-\omega$ contribution to the $NN$ interaction is constructed by
employing $NN-$meson vertices constrained by empirical two-nucleon data. The
$\rho-\omega$ mixing amplitude, on the other hand, is obtained from a
measurement of the pion form factor at the on-shell $\omega$-meson point
\cite{barkov85,miller90,coon87}.

	In one boson exchange (OBE) models of the NN force, nucleons
interact via the exchange of several mesons possessing different Lorentz
and isospin transformation properties~\cite{machl86,machl87}. In this
paper we are concerned with the mixing between the isoscalar
$\omega$ meson and the neutral member of the isovector $\rho$ meson.
The neutral $\omega$ meson couples in a minimal fashion to the conserved
baryon current
  \begin{equation}
    {\cal L}_{{\scriptscriptstyle NN}\omega}=
    g_{\omega}\bar{\psi}\gamma^{\mu}\psi{\omega}_{\mu} \;.
   \label{lomega}
  \end{equation}
The isovector $\rho$ meson, on the other hand, has a vector as well
as a tensor coupling to the nucleon
  \begin{equation}
    {\cal L}_{{\scriptscriptstyle NN}\rho}=
     g_{\rho}\bar{\psi}\gamma^{\mu}
     {\bf \tau}\cdot\psi{\bf \rho}_{\mu} +
     f_{\rho}\bar{\psi}\sigma^{\mu\nu}
     {\bf \tau}\cdot\psi{\partial_{\mu} \over 2M}{\bf \rho}_{\nu} \;.
   \label{lrho}
  \end{equation}
	Notice that the above definitions are standard except
for possible factors of two (some definitions include isospin
vertices as ${\bf 1}/2$ or ${\bf \tau}/2$). In addition, some
models include a $NN\omega$ tensor coupling. Since in this paper
we choose typical coupling constants determined by the Bonn group,
we use their conventions throughout this work~\cite{machl86,machl87}.

	Having constructed an interaction Lagrangian one can then
proceed to calculate the contribution from $\rho-\omega$ mixing to
the NN potential, one obtains (with
$\Gamma^{\mu}\equiv~i\sigma^{\scriptscriptstyle \mu\nu}
{q_{\nu}/2M}$ and $C_{\rho}\equiv f_{\rho}/g_{\rho}$)~\cite{henmil79,coon75}
 \begin{mathletters}
  \begin{eqnarray}
    \hat{V}^{\rho\omega}_{\scriptscriptstyle III}(q) &=&
      V^{\rho\omega}_{NN}(q) \,
       \gamma^{\mu}(1)\gamma_{\mu}(2)
       \Big[\tau_{\scriptscriptstyle z}(1)+
            \tau_{\scriptscriptstyle z}(2)\Big] \;,
     \label{vthree} \\
    \hat{V}^{\rho\omega}_{\scriptscriptstyle IV}(q) &=&
      V^{\rho\omega}_{NN}(q) \,
      C_{\rho}
      \Big[
        \Gamma^{\mu}(1)\gamma_{\mu}(2)
        \tau_{\scriptscriptstyle z}(1)-
        \gamma^{\mu}(1)\Gamma_{\mu}(2)
        \tau_{\scriptscriptstyle z}(2)
      \Big]  \;,
     \label{vfour}
  \end{eqnarray}
  \end{mathletters}
where
  \begin{equation}
     V^{\rho\omega}_{NN}(q)=-
     {g_{\rho}g_{\omega}
     \langle \rho | H | \,\omega \rangle \over
     (q^{2}-m_{\rho}^{2})
     (q^{2}-m_{\omega}^{2})}\;.
   \label{vrhoom}
  \end{equation}

This parameter-free construction of the potential is quite satisfactory
since it does not introduce additional parameters beyond those
already constrained by charge-symmetry-conserving (CSC) two-nucleon data.
More importantly perhaps, most of the differences observed experimentally
in the binding energy of mirror nuclei, the Nolen-Schiffer
anomaly~\cite{nolsch69}, have been attributed
to $\rho-\omega$ mixing and explained using the above
potential~\cite{bluiqb87}. In addition, $\rho-\omega$ mixing plays an
important role in explaining the difference between the
neutron and proton analyzing power ($\Delta A$) in elastic neutron-proton
scattering~\cite{abegg86,knutsen90,miller86,willia87}. Indeed,
$\rho-\omega$ mixing seems to account for half the size of the effect
in the IUCF experiment~\cite{miller90,knutsen90}.

      One important issue that has been overlooked until very
recently, however, is the momentum dependence of the $\rho-\omega$
mixing amplitude. To date, most of the theoretical efforts devoted to the
understanding of $\rho-\omega$ mixing in CSV have assumed the constant
on-shell value for the mixing amplitude. Since the relevant momentum
transfer carried by the meson exchange between two nucleons is spacelike
($q_{\mu}^{2}<0$), the use of a mixing amplitude determined at a timelike
point is clearly suspect. Recently, Goldman, Henderson, and Thomas have
tested the on-shell assumption by constructing a model in which the
mixing is driven by the $u-d$ quark mass difference~\cite{goldman91}.
Since in their model the mixing is generated by $q\bar{q}$ loops they
could examine the momentum dependence of the mixing amplitude and, thus,
confront the on-shell assumption. They have concluded that the on-shell
assumption may be surprisingly poor.

In this paper we study the momentum dependence of the $\rho-\omega$ mixing
amplitude in a purely hadronic model. We consider the mixing amplitude to
also be generated by fermion loops. The basic assumption of our model,
however, is that the mixing amplitude is generated by $N\bar{N}$ loops
and thus driven entirely by the small neutron-proton mass difference.

There are several advantages in calculating the $\rho-\omega$ mixing
amplitude using a purely hadronic description. First, all parameters used
in the calculation are known to great accuracy ({\it e.g.,} nucleon and
meson masses) or are constrained by empirical data ({\it e.g.,} coupling
constants)~\cite{machl86,machl87,walec74,serwal86}. Furthermore, since
field-theoretical models naturally include vacuum corrections, the coupling
between nucleons and antinucleons is determined by the underlying theory
and, thus, ultimately constrained by the empirical data. These vacuum
contributions are an essential part of the relativistic description of
a nuclear target. In fact, vacuum corrections are known to be crucial
for avoiding the appearance of spurious (center-of-mass) excitations
and for maintaining the conservation of the electromagnetic
current~\cite{piekar90,dawfur90}. Furthermore, hadronic models have
been relatively successful in calculating the nuclear response to
electromagnetic probes by including, in addition to the traditional
particle-hole excitations, vacuum polarization~\cite{horpie89,horpie90}.
The self-consistency of these calculations demanded that the same
interaction used in the calculation of ground state properties
be used for the residual particle-hole and $N\bar{N}$ interaction.
Hence, in the calculation of the $N\bar{N}$ loops driving
the $\rho-\omega$ mixing amplitude [see Eq.~({\ref{hrhoom}}) below],
we will use the same dynamical input as in the calculation of the
$\rho-\omega$ contribution to the NN potential [Eq.~({\ref{vrhoom}})].
Therefore, the $\rho-\omega$ mixing amplitude constitutes a parameter free
prediction. While one might question the reliability of a purely hadronic
description of the mixing amplitude, we feel that this is balanced by
the simplicity and parameter-free nature of the calculation.
Second, in the calculation of fermion loops there are no
unphysical ({\it e.g.,} $q\bar{q}$) thresholds appearing near the region
of interest~\cite{goldman91}. The only thresholds that appear are related
to the production of physical ($N\bar{N}$) states and occur far away from
the region of interest. Finally, there is no need to introduce ad-hoc
form factors to regularize (infinite) loop integrals (we discuss later
the sensitivity of our results to the introduction of form factors).
Since the mixing amplitude is sensitive only to the difference between
proton and neutron loops, the difference is finite even though the
individual pieces are not (note that the difference between u-quark and
d-quark loops should also be finite).

	Using standard Feynman rules, the $\rho-\omega$ mixing amplitude
in the hadronic model described by Eqs.~({\ref{lomega}})~and~({\ref{lrho}})
can be written as
  \begin{equation}
     \langle \rho | H | \,\omega \rangle =
     g_{\rho}\,g_{\omega}\,q^2\,\Pi(q^2) \;,
   \label{hrhoom}
  \end{equation}
where $\Pi(q^2)$ is the transverse component of the full polarization
tensor (see Eqs.~({\ref{pivv}})~and~({\ref{pivt}}) below). Because the
$\rho$ meson has, both, a vector and a tensor coupling to the nucleon,
one needs to evaluate polarization tensors having Lorentz vector and
tensor vertices, {\it i.e.,}
  \begin{equation}
     \Pi^{\mu\nu}(q) =
     \Pi^{\mu\nu}_{{\rm vv}}(q) + C_{\rho}
     \Pi^{\mu\nu}_{{\rm vt}} \;,
   \label{pimunu}
  \end{equation}
where
  \begin{mathletters}
   \begin{eqnarray}
     i\Pi^{\mu\nu}_{{\rm vv}}(q) &=&
     \int {d^4k\over (2\pi)^4} {\rm Tr}
     \left[
       \gamma^{\mu} G(k+q)
       \gamma^{\nu} \tau_{\scriptscriptstyle z} G(k)
      \right] \;,
      \label{pimunuvv}     \\
     i\Pi^{\mu\nu}_{{\rm vt}}(q) &=&
     \int {d^4k\over (2\pi)^4} {\rm Tr}
     \left[
       \gamma^{\mu} G(k+q)
       {i\sigma^{\scriptscriptstyle \nu\lambda}
       q_{\scriptscriptstyle \lambda}\over 2M}
       \tau_{\scriptscriptstyle z} G(k)
      \right] \;.
      \label{pimunuvt}
   \end{eqnarray}
  \end{mathletters}
The isospin trace can be evaluated by writing isoscalar and isovector
components of the nucleon propagator
  \begin{equation}
	G(k)={1\over 2}G_{p}(k)(1+\tau_{\scriptscriptstyle z}) +
	     {1\over 2}G_{n}(k)(1-\tau_{\scriptscriptstyle z}) \equiv
	     G_{0}(k)+G_{1}(k)\tau_{\scriptscriptstyle z}
  \end{equation}
in terms of individual proton and neutron contributions
  \begin{equation}
	G_{p}(k)=
         {\rlap/{k} + M_{p} \over
          k^2-M_{p}^{2}+i\epsilon} \;, \quad
	G_{n}(k)=
         {\rlap/{k} + M_{n} \over
          k^2-M_{n}^{2}+i\epsilon} \;.
   \label{greens}
  \end{equation}
After performing the isospin trace one discovers, as expected,
that the $\rho-\omega$ mixing amplitude is driven by the
difference between proton and neutron loops
  \begin{equation}
     \Pi_{\mu\nu}(q) = \Pi_{\mu\nu}^{(p)}(q) - \Pi_{\mu\nu}^{(n)}(q) \;.
   \label{pidiff}
  \end{equation}

	The calculation of the vacuum loops is now completely
standard~\cite{ramond81}. Since the individual proton and neutron
contribution diverge (but not their difference) we isolate the
singularities by using dimensional regularization. The Lorentz tensor
structure of the polarization can be obtained after performing the
appropriate traces, we obtain,
  \begin{mathletters}
   \begin{eqnarray}
     \Pi^{\mu\nu}_{{\rm vv}}(q) &=&
     \Big(
      -g^{\mu\nu}q^{2}+
       q^{\mu}q^{\nu}
     \Big)\Pi_{\rm vv}(q^2)  \;,
     \label{pivv}    \\
     \Pi^{\mu\nu}_{{\rm vt}}(q) &=&
     \Big(
      -g^{\mu\nu}q^{2}+
       q^{\mu}q^{\nu}
     \Big)\Pi_{\rm vt}(q^2) \;,
     \label{pivt}
   \end{eqnarray}
  \end{mathletters}
where the unrenormalized polarizations are given (for proton loops) by
  \begin{mathletters}
   \begin{eqnarray}
     \Pi^{(p)}_{{\rm vv}}(q^2) &=&
     -{1 \over 2\pi^2}
     \left[
      {1 \over 6\epsilon}-{\gamma \over 6}-
      \int_{0}^{1} dx \, x \, (1-x)
      \ln \left(M_{p}^{2}-x(1-x)q^{2} \over \Lambda^{2}\right)
     \right] \;,
     \label{pivvinf}    \\
     \Pi^{(p)}_{{\rm vt}}(q^2) &=&
     -{1 \over 8\pi^2}
     \left[
      {1 \over \epsilon}-\gamma-
      \int_{0}^{1} dx
      \ln \left(M_{p}^{2}-x(1-x)q^{2} \over \Lambda^{2}\right)
     \right] \;.
     \label{pivtinf}
   \end{eqnarray}
  \end{mathletters}
In the above equations $\Lambda$ is an arbitrary renormalization constant,
$\gamma$ is the Euler-Mascheroni constant, and $\epsilon \rightarrow 0$. A
finite mixing amplitude is now obtained by taking the difference
between proton and neutron contributions
  \begin{mathletters}
   \begin{eqnarray}
     \Pi_{{\rm vv}}(q^2) &=&
     \Pi_{{\rm vv}}^{(p)}(q^2)  -
     \Pi_{{\rm vv}}^{(n)}(q^2)  =
      {1 \over 2\pi^2}
      \int_{0}^{1} dx \, x \, (1-x)
      \ln \left[
        {M_{p}^{2}-x(1-x)q^{2} \over M_{n}^{2}-x(1-x)q^{2}}
     \right],
     \label{pivvfin}    \\
     \Pi_{{\rm vt}}(q^2) &=&
     \Pi_{{\rm vt}}^{(p)}(q^2)  -
     \Pi_{{\rm vt}}^{(n)}(q^2)  =
      {1 \over 8\pi^2}
      \int_{0}^{1} dx
      \ln \left[
        {M_{p}^{2}-x(1-x)q^{2} \over M_{n}^{2}-x(1-x)q^{2}}
     \right].
     \label{pivtfin}
   \end{eqnarray}
  \end{mathletters}
Expanding the integrand to first order in the neutron-proton mass
difference we obtain the following closed-form expression for
the $\rho-\omega$ mixing amplitude:
  \begin{eqnarray}
     {\langle \rho | H | \, \omega \rangle \over M^2} =
     {g_{\rho}g_{\omega} \over \pi^2}
     {\Delta M \over M}
     \cases{
       1-\displaystyle{1\over \xi}
       (1+\xi^2+C_{\rho})\,{\rm tan}^{-1}
       \left( \displaystyle {1 \over \xi} \right) \,, &
       for $0 < q^2 < 4M^2 \;;$ \cr
       ${\phantom {xxx}}$  &   \cr
       1-\displaystyle{1\over \xi}
       (1-\xi^2+C_{\rho}){1 \over 2}
       \ln \left( \left|
        \displaystyle{\xi-1 \over \xi+1}
       \right| \right) \,, &
       ${\rm otherwise} \;,$ \cr}
   \label{hrhooman}
  \end{eqnarray}
where
  \begin{equation}
   M = {1\over 2}(M_{n}+M_{p}) \;, \quad
   \Delta M= (M_{n}-M_{p})   \;, \ {\rm and} \quad
   \xi=\displaystyle{
    \left| 1-{4M^2 \over q^2} \right|^{1/2}} \;.
  \end{equation}
The above analytic expression is accurate everywhere except
in the vicinity of the thresholds ($q^{2}=4M^{2}$).
This equation embodies the central result of the present work.
It displays the momentum dependence of the $\rho-\omega$ mixing
amplitude in terms of three parameters ($g_{\omega}\,,g_{\rho},$ and
$C_{\rho}$). Having previously constrained these parameters from
CSC two-nucleon data, this result provides a parameter-free prediction
of the model. In particular, using the above expression at the on-shell
$\omega$-meson point, together with the parameters of
Table~{\ref{tableone}}, we obtain
  \begin{eqnarray*}
     {\langle \rho | H | \, \omega \rangle}
     \Big|_{q^2=m_{\scriptscriptstyle \omega}^{\scriptscriptstyle 2}}=
     -4314\,{\rm MeV}^{2} \;,
  \end{eqnarray*}
which compares well to the experimental value of
$(-4520\pm600)\,{\rm MeV}^{2}$.

	The momentum dependence of the $\rho-\omega$ mixing
amplitude is shown in Fig.~{\ref{figone}}. Two calculations
are displayed. The solid line shows results for the mixing
amplitude using Eq.~({\ref{hrhooman}}) at all values of
$q^{2}$. In contrast, the dashed line shows results modified
by the introduction of form factors in the spacelike region.
These form factors are introduced by modifying the point coupling
in the following way:
  \begin{equation}
    g_{\rho} \rightarrow  g_{\rho}(q^2) \equiv
    g_{\rho} \left(
    1-q^{2}/\Lambda_{\rho}^{2} \right)^{-1} \;; \quad
    g_{\omega} \rightarrow  g_{\omega}(q^2) \equiv
    g_{\omega} \left(
    1-q^{2}/\Lambda_{\omega}^{2} \right)^{-1} \;. \quad
   \label{formf}
  \end{equation}
As before, the numerical values for the cutoffs ($\Lambda$) are
constrained by empirical two-nucleon data (see Table~{\ref{tableone}}).

The topic of form factors, or vertex corrections, now needs some
comment. It is clear that the finite size of, both, nucleons and
mesons should modified the naive point coupling at finite $q^{2}$.
These vertex corrections, however, need not be included in an
ad-hoc fashion. In fact, vertex corrections can, in principle, be
calculated using renormalizable models based on hadronic degrees of
freedom. The basic dynamical assumption of these models is that the
internal structure of the hadrons can be described in terms of hadronic
degrees of freedom alone~\cite{serwal86}. Although some progress has
recently been made in calculating vertex corrections in hadronic
theories~\cite{allser92}, much work remains to be done. In particular,
very little has been said about vertex corrections in the timelike region.
In addition, OBE models of the NN interaction can only constrained the
form factors on-shell. Hence, in order to minimize the model assumptions
introduced in our calculation, we modify the naive point coupling only in
the spacelike region by the introduction of on-shell form factors as
prescribed by Eq.~({\ref{formf}}). The issue of off-shell form factors is
clearly an important and open problem.

Because of the smooth behavior of the (dimensionless) transverse
polarization over the sampled $q^{2}$ region (not shown), the most important
momentum behavior of the mixing amplitude displayed in Fig.~{\ref{figone}}
is determined by the $q^{2}$ factor in Eq.~({\ref{hrhoom}}). In particular,
it reveals that the mixing amplitude has the opposite sign relative to
its value at the on-shell point over the entire spacelike region sampled
in NN scattering. This behavior is clearly seen in Fig.~{\ref{figtwo}}
which shows the $\rho-\omega$ contribution to the NN potential. Three
calculations are displayed. The solid line shows the NN
potential~[Eq.~({\ref{vrhoom}})] obtained from using the on-shell
value for the $\rho-\omega$ mixing amplitude. The dashed line uses
the same on-shell amplitude but modifies the naive coupling by the
inclusion of on-shell form factors at the external nucleon legs. Finally,
the dash-dotted line uses the off-shell $\rho-\omega$ mixing amplitude
and has, both, the off-shell amplitude and the external nucleon legs
modified by form factors. In addition to the sign difference
displayed over the entire spacelike region, a much suppressed contribution
to the NN potential is observed whenever the off-shell mixing amplitude is
used.

Finally, in Fig.~{\ref{figthree}} we show the static contribution to the
NN potential in configuration space. These results are obtained from
taking the Fourier transform of the three momentum-space potentials
displayed in~Fig.~{\ref{figtwo}}. In particular, with the on-shell
mixing amplitude and with no on-shell form factors this $r$-space
potential takes the following form~\cite{henmil79,coon75}
  \begin{equation}
     V^{\rho\omega}_{NN}(r)=-
     {g_{\rho}g_{\omega} \over 4\pi}
     {\langle \rho | H | \,\omega \rangle \over
     m_{\omega}^{2}-m_{\rho}^{2}}
     \left(
      {e^{-m_{\rho}r}   \over r} -
      {e^{-m_{\omega}r} \over r}
     \right) \,.
  \end{equation}
Large differences are clearly seen in the
interior for all three potentials. However, due to the strong repulsive
nature of the charge-symmetry-conserving NN potential, the two-nucleon
wave function will be insensitive to the details of the distorting CSV
potential in the interior. More importantly, perhaps, is the occurrence
of a node in the potential around $r\sim 0.9$~fm. This is the region where
conventional estimates suggests that the $\rho-\omega$ contribution,
obtained from a competition between the fast fall-off of the CSV potential
and the suppression of the two-nucleon wave function in the interior,
should be larger. Consequently, our findings are in basic agreement with
the results obtained in Ref.~\cite{goldman91} and are consistent with
the view that the $\rho-\omega$ contribution to the CSV nucleon-nucleon
potential is, effectively, nonexistent.

	In summary, we have calculated the momentum dependence of the
$\rho-\omega$ mixing amplitude in a simple hadronic model. The mixing
was assumed to be generated solely by $N\bar{N}$ loops and thus driven
by the neutron-proton mass difference. Since the mixing is sensitive only
to the difference between proton and neutron loops, the amplitude
was found to be finite even though the individual pieces were
not. We have presented closed-form analytic expressions for the mixing
amplitude in terms of very few parameters. Furthermore, these parameters
were obtained from previous fits to two-nucleon data. Hence, our results
can be regarded as parameter-free predictions of one-boson exchange
models. Using standard values for these parameters we obtained a value
for the $\rho-\omega$ mixing amplitude at the on-shell $\omega$-meson
point in good agreement with experiment. We have extended our results
to the spacelike region and have computed the contribution from the
off-shell $\rho-\omega$ mixing amplitude to the NN potential. These results
were compared to a recent calculation of the mixing amplitude in terms of
$q\bar{q}$ loops\cite{goldman91}. In spite of the obvious differences
between the two models, our findings agree with the main conclusions
drawn from that work, namely, that the momentum dependence of the
$\rho-\omega$ mixing amplitude is significant and that the occurrence
of a node in the NN potential around $r\sim 0.9$~fm severly suppresses
the $\rho-\omega$ contribution to the CSV potential.

What will the
impact of these results be on CSV observables is, at this juncture,
hard to predict. For some of them like the Nolen-Schiffer anomaly or
the differences in NN scattering lengths one must, at the very least,
study the momentum dependence of the $\pi-\eta$ mixing
amplitude~\cite{maltman92}. In addition,
one must study medium modifications to the vector-meson propagators. This
effect might be important in explaining the Nolen-Schiffer anomaly
for medium- to heavy-nuclei. These issues are currently under investigation.

	The neutron-proton analyzing power difference, $\Delta A$, might,
however, pose a serious challenge. There, the $\pi-\eta$ amplitude does
not contribute. At $T_{\rm lab}=477$~MeV, $\Delta A$ is dominated by the
one-pion exchange potential and an absent $\rho-\omega$ contribution
might not spoil the agreement with experiment. At $T_{\rm lab}=183$~MeV,
on the other hand, the one-pion exchange contribution is small and
$\rho-\omega$ mixing dominates~\cite{miller90,miller86,willia87}.
Hence, if as suggested by our findings,
and by those of Ref.~\cite{goldman91}, $\rho-\omega$ mixing is indeed
severly suppressed, additional CSV mechanisms will have to be found in
order to account for the unexplained $\Delta A$ at $T_{\rm lab}=183$~MeV.

\acknowledgments
This research was supported by the Florida State University
Supercomputer Computations Research Institute and U.S. Department
of Energy contracts DE-FC05-85ER250000, DE-FG05-92ER40750,
and DE-FG05-86ER40273.

\figure{The $\rho-\omega$ mixing amplitude as a function of $q^{2}$
        with (dashed line) and without (solid line) the inclusion
        of form factors in the spacelike region. The experimental point
        was extracted from a measurement of the electromagnetic pion
        form factor at the $\omega$-meson point~\cite{barkov85}.
        \label{figone}}
\figure{The contribution from $\rho-\omega$ mixing to the NN potential
        as a function of $q^{2}$ using the off-shell value for the
        mixing amplitude (dash-dotted line), and the on-shell
        value with (dashed line) and without (solid line)
        the inclusion of on-shell form factors at the external
        nucleon legs.
        \label{figtwo}}
\figure{The contribution from $\rho-\omega$ mixing to the NN potential
        as a function of the NN separation using the off-shell value
        for the mixing amplitude (dash-dotted line), and the on-shell
        value with (dashed line) and without (solid line)
        the inclusion of on-shell form factors at the external
        nucleon legs.
        \label{figthree}}

 \mediumtext
 \begin{table}
  \caption{Meson masses, coupling constants, tensor-to-vector ratio
           and cutoff parameters in the Bonn one-boson exchange model
           (see Table~4.2 of Ref.~\cite{machl86} and
            Table~4 of Ref.~\cite{machl87}).}
   \begin{tabular}{ccccc}
    Meson & Mass(MeV) & ${g^2/4\pi}$
          & $C=f/g$ & $\Lambda({\rm MeV})$ \\
        \tableline
    $\rho$    &  770  &  0.41 &  6.1  &  1400  \\
    $\omega$  &  783  &  10.6 &  0.0  &  1500  \\
   \end{tabular}
  \label{tableone}
 \end{table}

\end{document}